\documentclass[fleqn,twoside]{article}
\usepackage{espcrc2}
\usepackage{graphicx}
\usepackage[figuresright]{rotating}

\def\d{\hbox{d}} 

\title{Infrared structure of $e^+e^- \to 3$~jets at NNLO - the $C_F^2$
contribution}

\author{A.~Gehrmann--De Ridder\address[Zu]{Institut f\"ur Theoretische Physik, Universit\"at Z\"urich,
Winterthurerstrasse 190, CH-8057 Z\"urich, Switzerland}%
        ,
        T.~Gehrmann\addressmark[Zu] and
        E.W.N.~Glover\address{Institute for Particle Physics Phenomenology, University of Durham,
South Road, Durham DH1 3LE, England}
        }
       
\begin{document}

\begin{abstract}
We discuss the infrared structure of $e^+e^- \to 2$ and 3~jets at NNLO in QCD
perturbation theory and describe subtraction terms that render the separate
parton-level contributions finite.  As a first result, we find that the NNLO
$C_F^2$ contribution to the first moment of the Thrust distribution 
$\langle 1-T \rangle = -20.4 \pm 4$.  
\vspace{1pc}
\end{abstract}

\maketitle

\section{Introduction}

Hadronic event shapes and jet production observables can be 
measured very accurately at LEP and future high
energy electron positron colliders. 
By confronting these data with theoretical calculations, one
can determine the strong coupling constant.  Analyzing the different sources of error on these
determinations, it becomes clear that the largest source of uncertainty 
is theoretical and mainly due to the truncation of the perturbation series
at next-to-leading order (NLO). 

To improve this situation, the calculation of next-to-next-to-leading order
(NNLO) corrections to jet observables  is mandatory. 
For an $n$-jet observable,
several ingredients are required; the two-loop $n$-parton matrix elements,  the
one-loop ($n$+1)-parton matrix elements and the tree level ($n$+2)-parton 
matrix elements.

In the specific case of three-jet final states in 
$e^+e^-$ annihilation   
the primary process is
$\gamma^{*} \to q \bar{q} g$, the decay of a virtual photon into a
 quark--antiquark
pair accompanied by a gluon.
The individual  partonic channels that contribute through to NNLO 
are shown in Table~1.
\begin{table}[h]
\caption{The partonic channels contributing to $e^+e^- \to 3$~jets.}
\begin{tabular}{lll}
\hline\\
LO & $\gamma^*\to q\,\bar qg$ & tree level \\[2mm]
NLO & $\gamma^*\to q\,\bar qg$ & one loop \\
 & $\gamma^*\to q\,\bar q\, gg$ & tree level \\
 & $\gamma^*\to q\,\bar q\, q\bar q$ & tree level \\[2mm]
NNLO & $\gamma^*\to q\,\bar qg$ & two loop \\
 & $\gamma^*\to q\,\bar q\, gg$ & one loop \\
& $\gamma^*\to q\,\bar q\, q\,\bar q$ & one loop \\
& $\gamma^*\to q\,\bar q\, q\,\bar q\, g$ & tree level \\
& $\gamma^*\to q\,\bar q\, g\,g\,g$ & tree level\\[2mm]
\hline
\vspace{-1cm}
\end{tabular}
\end{table}
In the recent past, enormous progress has been made  in the
calculation of two-loop QCD matrix elements, which are
now known for  
$\gamma^*\to q\bar q g$~\cite{3jme,muw2}.  The one-loop matrix elements 
with one additional parton~\cite{V4p}  and the tree-level matrix elements with two
more partons  are also known and form part of NLO programs for 
four-jet production~\cite{nplusone1,cullen}.

\section{The infrared problem}

Now that all the pieces are available, all that remains is to combine them in a
way that produces numerically stable results for physical observables. To
achieve this, the contributions of the processes shown in Table~1  are weighted
by  jet functions  which select three-jet final states from the partonic final
state momenta. At a given  order, all partonic multiplicity channels
contributing to this order  have to be summed. However, each partonic channel
contains infrared singularities which, after summation, cancel among each
other. While infrared singularities from purely virtual corrections are
obtained  immediately after integration over the loop momenta, their extraction
is  more involved for real emission (or mixed real-virtual) contributions.
Here, the infrared singularities become only explicit after integrating  the
real radiation matrix elements over the phase space appropriate to  the jet
observable under consideration. 

Exactly how to accomplish this task is presently an open question - see
Refs.~\cite{babistalk,gudruntalk,stefantalk} and the problem of integrating out
double real emission contributions  has so far only been addressed in specific 
calculations~\cite{ggam,uwer,fg,daleo}, each of which  requires a subset of the
ingredients needed for generic jet observables at NNLO.

The infrared singularities   of the real radiation contributions can be
extracted using  infrared subtraction  terms. These subtraction terms are
constructed such that they approximate the full  real radiation matrix elements
in all singular limits while still being  sufficiently simple to be integrated
analytically over a section of phase space that encompasses all regions
corresponding to singular configurations.  In all cases, the subtraction terms
must be local in phase space.   However, there are two distinct approaches to
derive the subtraction terms.   In the first~\cite{babistalk,gudruntalk}, one
uses phase space remappings  together with the iterated  sector
decomposition~\cite{hepp,itsec} to extract the singularities from individual
terms in the matrix elements in terms of plus 
prescriptions~\cite{itsec1,ggh,babis,babisnew}. In the
second~\cite{stefantalk}, one identifies one- and two-particle subtraction
functions that approximate the full matrix elements in all of the singular
limits and which are sufficiently simple to be integrated analytically over the
unresolved phase space.  It is this latter approach that we follow here.

One-particle subtraction at tree level is well understood from NLO 
calculations  and general algorithms are available for  one-particle
subtraction at one-loop~\cite{onel,cg,kos1,weinzierl2}, in a  form that has
recently been integrated  analytically~\cite{kos1,weinzierl2}.

Similarly, tree-level two-particle subtraction terms have been extensively
studied in the  literature~\cite{twot,kosower,weinzierl1}. However  their
integration over the unresolved phase space remains an outstanding  issue.

To specify the notation, we define 
the tree level $m$-parton contribution to the $J$-jet cross section 
in $d$-dimensions by,
\begin{equation}
{\rm d} \sigma^{B}\sim {\rm d}\Phi_{m}
\,|{\cal M}_{m}|^{2}\; 
{\cal F}_{J}^{(m)}.
\label{eq:sigm}
\end{equation}
where $\sim$ hides all QCD-independent factors,
the sum over all configurations with $m$ partons and symmetry
factors for identical partons in the final state.
${\rm d}\Phi_{m}$ is the phase space for $m$ partons
and $|{\cal M}_{m}|$ is the tree level $m$-parton matrix element.
The jet function $ {\cal F}_{J}^{(m)}$ defines the procedure for 
building $J$-jets out of $m$ partons.
The main property of ${\cal F}_{J}^{(m)}$ is that the jet observable defined
above is collinear and infrared safe.

\section{NLO infrared subtraction terms}

At NLO, we consider the following $m$-jet cross section,
\begin{eqnarray}
{\rm d}\sigma^m_{NLO}&=&\int_{{\rm d}\Phi_{m+1}}\left({\rm d}\sigma^{R}_{NLO}
-{\rm d}\sigma^{S}_{NLO}\right) \\
&+&\left [\int_{{\rm d}\Phi_{m+1}}
{\rm d}\sigma^{S}_{NLO}+\int_{{\rm d}\Phi_{m}}{\rm d}\sigma^{V}_{NLO}\right]. \nonumber
\end{eqnarray}
The cross section ${\rm d}\sigma^{R}_{NLO}$ has the same expression as the 
Born cross section ${\rm d}\sigma^{B}_{NLO}$ above
except that $m \to m+1$, while 
${\rm d}\sigma^{V}_{NLO}$ is the one-loop virtual correction to the 
$m$-parton Born cross section ${\rm d}\sigma^{B}$.
The cross section ${\rm d}\sigma^{S}_{NLO}$ is a local counter-term for  
 ${\rm d}\sigma^{R}_{NLO}$. It has the same unintegrated
singular behavior as ${\rm d}\sigma^{R}_{NLO}$ in all appropriate limits.
Their difference is free of divergences 
and can be integrated over the $(m+1)$-parton phase space numerically.
The subtraction term  ${\rm d}\sigma^{S}_{NLO}$ has 
to be integrated analytically over all singular regions of the 
$m+1$-parton phase space. 
The resulting cross section added to the virtual contribution 
yields an infrared finite result. 

A systematic procedure for finding NLO infrared subtraction terms in the 
second method is the dipole formalism 
derived by Catani and Seymour~\cite{cs}.
Their subtraction terms are obtained as 
 sum of dipoles $\sum {\cal D}_{ijk}$ (where each dipole corresponds to a
single infrared singular configuration)
such that,
\begin{eqnarray}
\lefteqn{{\rm d}\sigma_{NLO}^{R}-{\rm d}\sigma_{NLO}^{S}
= {\rm d}\Phi_{m+1} 
\Bigg [|{\cal M}_{m+1}|^{2}\;
{\cal F}_{J}^{(m+1)} }\nonumber \\
\lefteqn{-\sum_{{\rm pairs}\, i,j}\!\sum_{k \neq i,j}{\cal D}_{ijk}\!
|{\cal M}_{m}((\tilde{p}_{ij},\tilde{p}_{k})|^2\,
{\cal F}_{J}^{(m)}(\tilde{p}_{ij},\tilde{p}_{k})\;\Bigg ].}\nonumber 
\label{eq:sub1}
\end{eqnarray}
The dipole contribution ${\cal D}_{ijk}$ involves the $m$-parton amplitude 
depending on the redefined on-shell momenta
$\tilde{p}_{ij},\tilde{p}_{k}$ 
and the dipole ${\cal D}_{ijk}$ which depends only on 
$p_{i},p_{j},{p}_{k}$.
The momenta ${p}_{i}$, $p_j$ and ${p}_{k}$ are respectively 
the emitter, unresolved parton  and the spectator momenta 
corresponding to a single dipole term. The redefined on-shell momenta 
$\tilde{p}_{ij},\tilde{p}_{k}$ are linear combinations of them.

\section{NNLO infrared subtraction terms}
\label{sec:nnlosub}
At NNLO, $m$-jet production is induced by final states containing up to
$(m+2)$ partons, including the one-loop virtual corrections to 
$(m+1)$-parton final 
states. As at NLO, one has to introduce subtraction terms for the 
$(m+1)$- and $(m+2)$-parton contributions. 
Schematically the NNLO $m$-jet cross section reads,
\begin{eqnarray}
{\rm d}\sigma^m_{NNLO}&=&\int_{{\rm d}\Phi_{m+2}}\left({\rm d}\sigma^{R}_{NNLO}
-{\rm d}\sigma^{S}_{NNLO}\right) \nonumber \\
\lefteqn{+\int_{{\rm d}\Phi_{m+1}}\left({\rm d}\sigma^{V,1}_{NNLO}
-{\rm d}\sigma^{VS,1}_{NNLO}\right)}\nonumber \\
\lefteqn{ +\int_{{\rm d}\Phi_{m+2}} {\rm d}\sigma^{S}_{NNLO}
+\int_{{\rm d}\Phi_{m+1}}{\rm d}\sigma^{VS,1}_{NNLO} }\nonumber\\
\lefteqn{+ \int_{{\rm d}\Phi_{m}}{\rm d}\sigma^{V,2}_{NNLO}\;,}  
\end{eqnarray}
where $\d \sigma^{S}_{NNLO}$ denotes the real radiation subtraction term 
coinciding with the $(m+2)$-parton tree level cross section  $\d
\sigma^{R}_{NNLO}$ in all singular limits.  Likewise, $\d \sigma^{VS,1}_{NNLO}$
is the one-loop virtual subtraction term  coinciding with the one-loop
$(m+1)$-parton cross section  $\d \sigma^{V,1}_{NNLO}$ in all singular limits. 
Finally, the two-loop correction  to the $m$-parton cross section is denoted by
${\rm d}\sigma^{V,2}_{NNLO}$.

In the simple case of two-jet production, 
the subtraction terms have been fully
identified~\cite{DGGG}. 
The four-particle contribution is
\begin{equation}
{\rm d}\sigma^{R}_{NNLO}
={\rm d}\Phi_{4}|{\cal M}_{4}|^{2} {\cal F}_{2}^{(4)}.
\end{equation}
Motivated by the fact that for three-jet production,
the sum over dipoles is essentially equivalent to the three-parton matrix element, 
the four-parton subtraction term is given by,
\begin{eqnarray}
{\rm d}\sigma^{S}_{NNLO}
&=&{\rm d}\Phi_{4}\left[|{\cal M}_{4}|^{2}
{\cal F}_{2}^{(2)}\right.\nonumber\\
\lefteqn{ +
\left.\sum_{ijk} |{\cal M}_{3}|^{2} D_{ijk} 
\left({\cal F}_{2}^{(3)}-{\cal F}_{2}^{(2)}\right) 
\right]. }
\end{eqnarray}
The first term precisely cancels the real radiation when the four-parton
configuration is perceived as a two-jet event.   The second removes
configurations when only one parton is unresolved, but the jet algorithm still
sees only two jets.   The final term eliminates cases when we double count.
Taken together, Eqs. 5 and 6 yield a finite result.
The one-loop three-parton contribution is,
\begin{equation}
{\rm d}\sigma^{V,1}_{NNLO}
={\rm d}\Phi_{3}|{\cal M}_{3}^{V,1}|^{2} {\cal F}_{2}^{(3)}
\end{equation}
and the subtraction term is
\begin{eqnarray}
{\rm d}\sigma^{VS,1}_{NNLO}
&=&{\rm d}\Phi_{3}\left[|{\cal M}_{3}^{V,1}|^{2}
{\cal F}_{2}^{(2)}\right.\nonumber\\
\lefteqn{ -
\left.\sum_{ijk} |{\cal M}_{3}|^{2} D_{ijk} 
\left({\cal F}_{2}^{(3)}-{\cal F}_{2}^{(2)}\right) 
\right]. }
\end{eqnarray}
As before, we subtract the full virtual matrix elements
so that the first term precisely cancels ${\rm d}\sigma^{V,1}_{NNLO}$
when the three-parton
configuration is perceived as a two-jet event.   The second and third
terms are merely the reappearance of the four-parton subtraction terms when a
single parton is unresolved.
Taken together, Eqs. 7 and 8 yield a finite result.
Finally, the two-parton contribution is made up of the two-loop contribution
\begin{equation}
{\rm d}\sigma^{V,2}_{NNLO} = {\rm d}\Phi_{2}|{\cal M}_{2}|^2  
|M^{V,2}_2|^2{\cal F}_{2}^{(2)} 
\end{equation}
and the analytically integrated subtractions terms (the first term on the RHS
of Eqs. 6 and 8),
\begin{eqnarray}
\lefteqn{\int_{{\rm d}\Phi_{m+2}} {\rm d}\sigma^{S}_{NNLO}
+\int_{{\rm d}\Phi_{m+1}}{\rm d}\sigma^{VS,1}_{NNLO} =}\nonumber \\
\lefteqn{{\rm d}\Phi_{2}|{\cal M}_{2}|^2 \left[ 
{\int_{\d \Phi_T} \!\!\!\!|M_4|^2} + \!\!\!{\int_{\d \Phi_D}\!\!\! 
|M_3^{V,1}|^2 }\right]  {\cal F}_{2}^{(2)}}
\end{eqnarray}
where,
the matrix elements are normalized to the two-parton matrix element 
such that
\begin{equation}
|M_{j}|^2 \equiv \frac{1}{|{\cal M}_{2}|^2} \, |{\cal M}_{j}|^2\;.
\end{equation}
and
\begin{equation}
{\rm d}\Phi_{3}= {\rm d}\Phi_{2}\; {\rm d}\Phi_{D},
\qquad{\rm d}\Phi_{4} =  {\rm d}\Phi_{2}\,{\rm d}\Phi_{T},
\end{equation}
defines the dipole and tripole phase space, ${\rm d}\Phi_{D}$ and
${\rm d}\Phi_{T}$ that the subtraction terms must be integrated over.
All of the master integrals necessary to perform the integrations in Eq.~10
have been evaluated in Ref.~\cite{ggh}. The result is an analytic expression 
in $d$-dimensions such that when taken together, Eqs. 9 and 10 yield a finite result.
For the inclusive hadronic $R$-ratio, ${\cal F}_J^{(m)}=1$ and the only
remaining contribution is the two-parton piece which agrees with that found in
the literature~\cite{chetyrkin}.
Similar results for distributions have been found using the sector
decomposition method~\cite{babisnew}.

\section{Three-jet event shapes at NNLO}

The subtraction terms based on subtracting the full matrix element  described
above are specific to the two-jet configurations.   Subtracting the full
tree-level  five-parton matrix elements and integrating it analytically is
neither sensible nor feasible.   On the other hand, one might expect that the
singular behaviour of the five-parton matrix elements can be represented by
simpler building blocks that depend on only four of the five parton momenta
multiplied by three-parton matrix elements that depend on momenta built from the
original parton momenta.   In this case, we can repeat the same steps as for
the two-jet subtraction terms and use Ref.~\cite{ggh} to analytically
integrate the subtraction term. 

At NNLO, the three-jet cross section contains seven colour structures,
\begin{equation}
\frac{1}{\sigma_0}{\rm d}\sigma^3_{NNLO} =
C_F \left[ A C_A^2 + B C_A C_F + C C_F^2 \right.
\end{equation}
$$
\left. +\! D C_A N_F + \!E C_F N_F + \!F N_F^2 +\! G
N_{F,\gamma}\left(\frac{4}{N}\!-\!N\right)\right].\nonumber 
$$
Because of the QED-like behaviour of the $C_F^2$ colour factor contribution,
the subtraction terms (for this colour factor) are very similar to the two-jet
case.   Accordingly,  we have implemented these terms in the NLO four-jet
program {\tt EERAD2}~\cite{cullen} and find that the five- and four-parton
contributions are numerically finite.   At the same time, the analytic
integration of the subtraction terms precisely cancels the infrared poles
found in the two-loop matrix elements~\cite{3jme}.

\subsection{The average value of 1-Thrust}
One of the classic event shape variables is Thrust. The first moment of the
Thrust distribution  $\langle 1-T\rangle$ is safe from large infrared logarithms and is therefore 
a theoretically clean and experimentally relevant observable.
The perturbative expansion of $\langle 1-T\rangle$ is given by
\begin{eqnarray*}
\langle 1-T \rangle &=& \int (1-T)\frac{1}{\sigma_0}\frac{d\sigma}{dT} \\
\lefteqn{\phantom{~} \hspace{-1cm}=C_F \left[ \left(\frac{\alpha_s}{2\pi}\right)A + \left(\frac{\alpha_s}{2\pi}\right)^2 B
+\left(\frac{\alpha_s}{2\pi}\right)^3 C + \ldots\right] }
\end{eqnarray*}
where $A = 1.57, B = 32.3$.  The NNLO coefficient $C$ receives contributions
from all seven colour structures.   We find that the $C_F^2$ colour
factor has the value,
$$C|_{C_F^2} = -20.4 \pm 4.$$

\section{Summary}
We have discussed the infrared singularity structure of jets in
electron-positron annihilation at NNLO.   For 2-jet events the subtraction
terms necessary to render the individual partonic contributions finite has been
worked through in detail~\cite{DGGG}.    In the case of 3-jet events, the
subtraction terms for the $C_F^2$ colour factor have been identified and
analytically integrated.   We have implemented these in a numerical program and
produced first NNLO results for a 3-jet event shape.
Analogous results for the remaining six colour structures are in progress.

\section*{Acknowledgments}
This research was supported in part by the Swiss National Funds 
(SNF) under contract 200021-101874, 
by PPARC and by
the EU Fifth Framework Programme  `Improving Human Potential', Research
Training Network `Particle Physics Phenomenology  at High Energy Colliders',
contract HPRN-CT-2000-00149.

\end{document}